\documentclass[a4paper,11pt]{article}
\pdfoutput=1 

\usepackage{jcappub} 

\usepackage{hyperref}

\usepackage[T1]{fontenc} 
\usepackage{footnote}
\def\be{\begin{equation}}
\def\ee{\end{equation}}
\def\ba{\begin{eqnarray}}
\def\ea{\end{eqnarray}}

\title{\boldmath{CMB and BAO constraints for an induced gravity dark energy model with a quartic potential}}

\author[a,b,c]{C. Umilt\`a}
\author[d,e,f]{M. Ballardini}
\author[e,f]{F. Finelli}
\author[e,f]{and D. Paoletti}
\affiliation[a]{Institut d'Astrophysique de Paris, CNRS (UMR7095), 98 bis Boulevard
Arago, F-75014, Paris, France}
\affiliation[b]{UPMC Univ Paris 06, UMR7095, 98 bis Boulevard Arago, F-75014, Paris,
France}
\affiliation[c]{Sorbonne Universit\'es, Institut Lagrange de Paris (ILP), 98 bis Boulevard
Arago, 75014 Paris, France}
\affiliation[d]{DIFA, Dipartimento di Fisica e Astronomia, Via Berti Pichat, I-40129 Bologna, Italy}
\affiliation[e]{INAF-IASF Bologna, via Gobetti 101, I-40129 Bologna, Italy}
\affiliation[f]{INFN, Sezione di Bologna, Via Irnerio 46, I-40126 Bologna, Italy}

\emailAdd{umilta@iap.fr}
\emailAdd{ballardini@iasfbo.inaf.it}
\emailAdd{finelli@iasfbo.inaf.it}
\emailAdd{paoletti@iasfbo.inaf.it}

\abstract{We study the predictions for structure formation 
in an induced gravity dark energy model with a quartic potential. By developing a dedicated 
Einstein-Boltzmann code, we study self-consistently the dynamics of homogeneous cosmology 
and of linear perturbations without using any parametrization. 
By evolving linear perturbations with initial conditions in the radiation era, we accurately recover the quasi-static 
analytic approximation in the matter dominated era. 
We use {\sc Planck} 2013 data and a compilation of baryonic acoustic oscillation (BAO) data
to constrain 
the coupling $\gamma$ to the Ricci curvature and the other cosmological parameters. 
By connecting the gravitational constant in the Einstein equation to the one 
measured in a Cavendish-like experiment, we find $\gamma < 0.0012$ at 95\% CL with {\sc Planck} 2013 and BAO data. 
This is the tightest cosmological constraint on $\gamma$ and on the corresponding derived 
post-Newtonian parameters. 
Because of a degeneracy between $\gamma$ and the Hubble constant $H_0$, we show how larger values for $\gamma$ are allowed, but not 
preferred at a significant statistical level, 
when local measurements of $H_0$ are combined in the analysis with {\sc Planck} 2013 data.
}

\begin{document}
\maketitle
\flushbottom

\section{Introduction}
\label{sec:intro}

Inflation or quintessence are naturally embedded in scalar-tensor theories of gravity.
In these models the scalar field which regulates the gravitational coupling also drives the 
acceleration of the Universe. The non-minimal coupling to gravity can change significantly 
the perspective on inflation or quintessence in Einstein general relativity. 
In the inflationary context, for instance, a large coupling of the inflaton 
to gravity allows potentials with a self-coupling which would be excluded 
in the minimally coupled case \cite{Spokoiny:1984bd,Lucchin:1985ip,Salopek:1988qh,Fakir:1990eg}.
In the dark energy context, for example the coupling to gravity could allow super-acceleration 
(i.e., $\dot H > 0$) with standard kinetic terms for the 
scalar field \cite{Boisseau:2000pr}. 


In this paper we consider
{\em induced gravity} (IG) with a quartic 
potential $V (\sigma) = \lambda \sigma^4/4$ as a simple scalar-tensor dark energy model:
\begin{equation}
S = \int d^4x \sqrt{-g}\, \Bigl[ \frac{\gamma \sigma^2 R}{2} - \frac{g^{\mu \nu}}{2}
\partial_{\mu} \sigma \partial_{\nu} \sigma - \frac{\lambda}{4} \sigma^4 + \mathcal{L}_m \Bigr] \,.
\label{IGself}
\end{equation}
where $\mathcal{L}_m$ denotes the contribution by matter and radiation.
Under a simple field redefinition $\gamma \sigma^2 = \phi/(8 \pi)$, 
the above action can be cast in a Brans-Dicke-like model \cite{BD} with a quadratic potential:
\begin{equation}
S= \int d^4x \sqrt{-g}\, \Bigl[\frac{1}{16 \pi} 
\left( \phi R  - \frac{\omega_\mathrm{BD}}{\phi} g^{\mu \nu} \partial_{\mu} \phi \partial_{\nu} \phi \right) - 
\frac{m^2}{2} \phi^2
+ \mathcal{L}_m \Bigr],
\end{equation}
with the following relation between the dimensionless parameters of the two theories:
\begin{equation}
\omega_\mathrm{BD} = \frac{1}{4 \gamma} \, \quad m = \frac{\sqrt{2 \lambda}}{16 \pi \gamma} .
\end{equation}

The action in Eq.~(\ref{IGself}), which contains only dimensionless parameters, was introduced to generate 
the gravitational constant and inflation by spontaneous breaking of scale invariance in absence of matter \cite{zee}.
In the context of late cosmology, this action was studied in Refs.~\cite{CV, Wetterich:1987fm} to reduce the time dependence of the effective gravitational constant 
in the original Brans-Dicke model (i.e., with a vanishing potential \cite{BD}) and to generate an 
effective cosmological constant. 
The cosmological background
dynamics from Eq.~(\ref{IGself}) was shown to be consistent with observations for small $\gamma$, i.e., $\gamma \lesssim 10^{-2}$ \cite{FTV}. 

The potential term in Eq.~(\ref{IGself}) is important for the global dynamics of the model and modifies 
the original Brans-Dicke attractor with power-law time dependence
of the scalar field in presence of non-relativistic matter, 
i.e., $a (t) = (t/t_0)^{(2 \omega_\mathrm{BD} + 2)/(3 \omega_\mathrm{BD} +4)}$ and 
$\Phi = \Phi_0 (t/t_0)^{2/(3 \omega_\mathrm{BD} +4)}$. 
At recent times, the potential term drives the Universe into acceleration and Einstein gravity plus a cosmological 
constant with a time-independent value of the scalar field emerge as an attractor among 
homogeneous cosmologies for the model in Eq.~(\ref{IGself}).

In this paper we study structure formation in the IG dark energy with a quartic potential in Eq.~(\ref{IGself}).
We study how gravitational instability at linear level 
depends on $\gamma$ through a dedicated Einstein-Boltzmann code.
We then use these theoretical predictions for cosmological observables to constrain the model with the 
{\sc Planck} 2013 data \cite{Ade:2013ktc,Ade:2013kta,Ade:2013zuv}, 
a compilation of baryonic acoustic oscillations (BAO) 
data \cite{Beutler:2011hx,Ross:2014qpa,Anderson:2013zyy} and local measurements of the 
Hubble constant \cite{Riess:2011yx,Humphreys:2013eja,Efstathiou:2013via}. 

\section{Dark Energy within Induced Gravity}
\label{sec:background}


The Friedmann and the Klein-Gordon equations for IG in a flat Robertson-Walker metric
are respectively:
\begin{equation}\label{fried-ig}
H^2 + 2 H \frac{\dot \sigma}{\sigma} = \frac{\sum_i \rho_i +V(\sigma)}{3\gamma \sigma^2} 
+ \frac{{\dot \sigma}^2}{6 \gamma \sigma^2} 
\end{equation}
\begin{equation}\label{kg-ig}
\ddot{\sigma} + 3H{\dot \sigma}+\frac{{\dot \sigma}^2}{\sigma} 
+ \frac{1}{(1+6\gamma)} \Bigl( V_{,\sigma}-\frac{4V}{\sigma}\Bigr) 
= \frac{1}{(1+6\gamma)} \frac{\sum_i(\rho_i -3p_i)}{\sigma}
\end{equation}
once the Einstein trace equation: 
\begin{equation}
- \gamma \sigma^2 R = T - (1+6 \gamma) \partial_\mu \sigma \partial^\mu \sigma 
- 4 V - 6 \gamma \sigma \Box \sigma
\end{equation}
is used. In the above $V_{,\sigma}$ denotes the derivative of the potential $V(\sigma)$ with respect 
to $\sigma$, the index $i$ runs over all fluid components, i.e. baryons, cold dark matter (CDM), photons and neutrinos, and we use a dot for the 
derivative with respect to the cosmic time. 
When considering $V \propto \sigma^4$ the potential cancels out from the Klein-Gordon equation and 
the scalar field is driven by non-relativistic matter. 
In the rest of the paper we will restrict ourselves to $V(\sigma) = \lambda \sigma^4/4$. 

\begin{figure}[t!!]
\centering
\epsfig{file=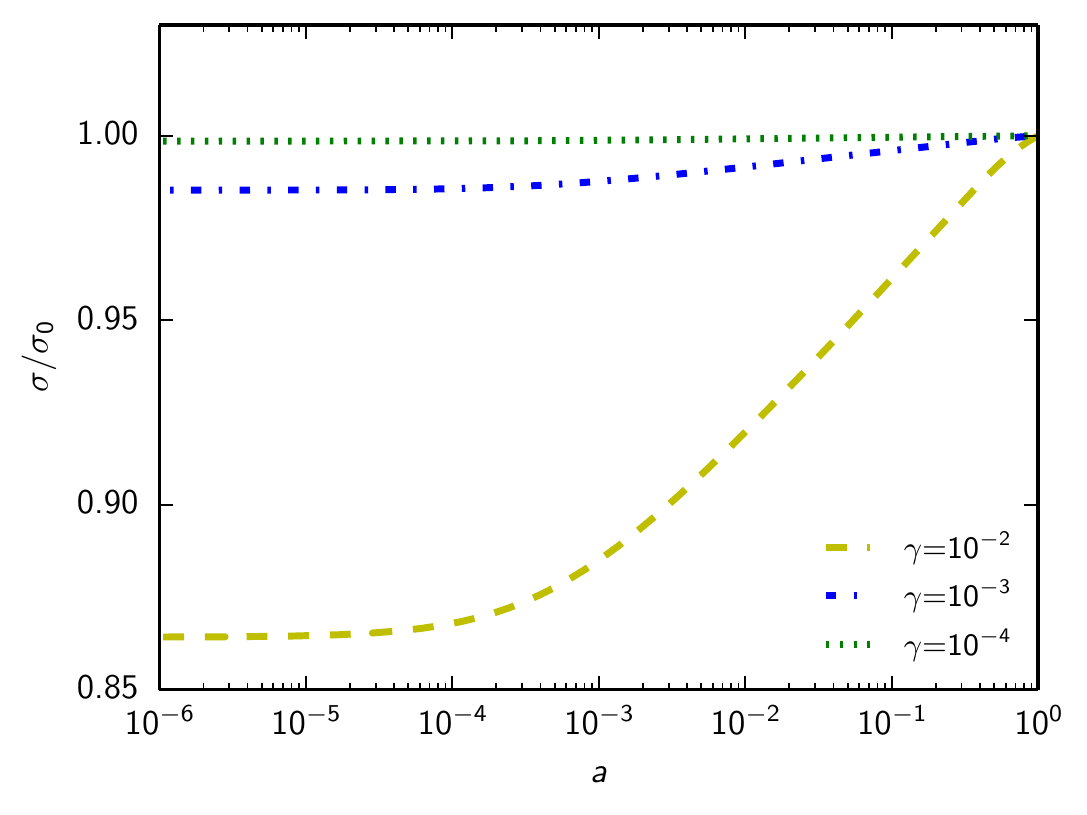, width=5 cm} \epsfig{file=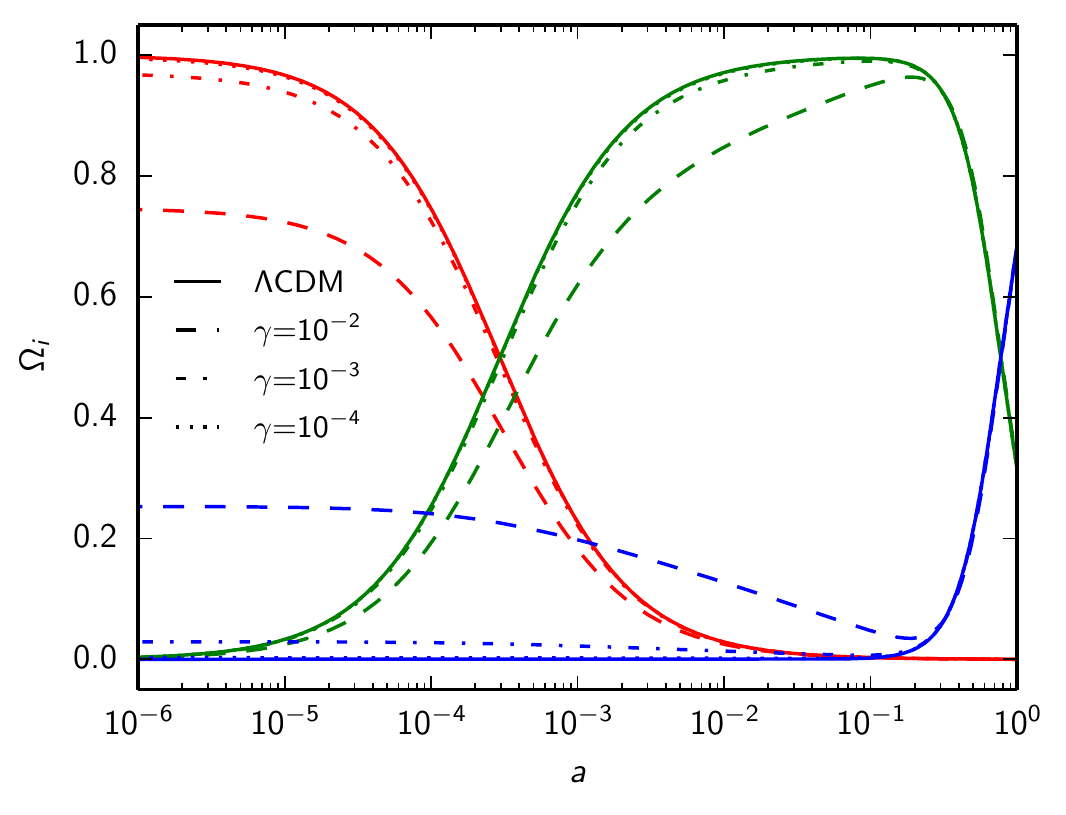, width=5 cm} \epsfig{file=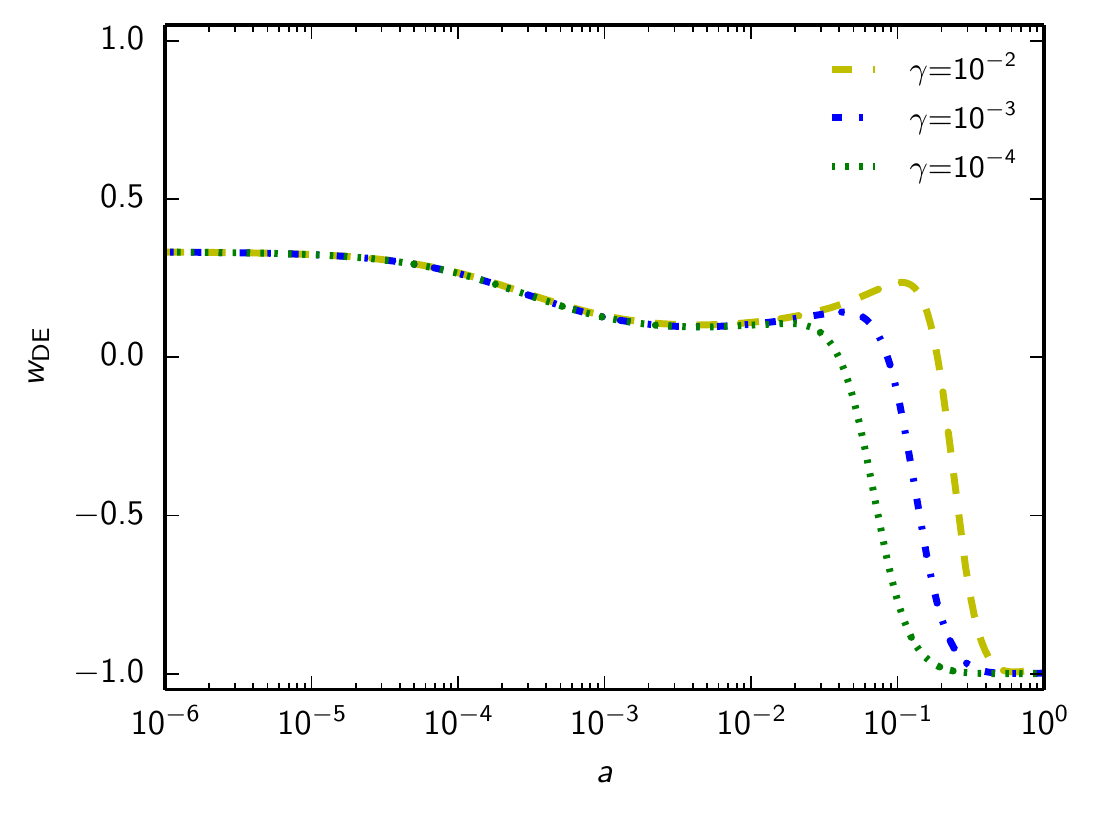, width=5 cm}
\caption{Evolution of $\sigma/\sigma_0$ (left panel), $\Omega_i$ (middle panel) and $w_{\rm DE}$ (right panel) as function of $\ln (a)$
for different choices of $\gamma$.}
\label{phiws}
\end{figure}

We consider the scalar field $\sigma$ at rest deep in the radiation era, since an initial non-vanishing time derivative would be 
rapidly dissipated \cite{FTV}. The scalar field is then driven by non-relativistic matter to an
asymptotically value higher than the one it had in the radiation era as can be seen in the left panel of Fig.~(\ref{phiws});    
when the scalar field freezes the Universe is driven in a de Sitter era by the scalar field potential 
which behaves as an effective cosmological constant \cite{CV,FTV}, as can be seen in the central panel of Fig.~(\ref{phiws}). 
Since $\sigma$ regulates the gravitational strength in the Friedmann equations, the present value of the field 
$\sigma_0$ can be connected the gravitational constant 
$G$ measured in laboratory Cavendish-type and solar system experiments by the relation: 
\begin{equation}
\gamma \sigma_0^2 = \frac{1}{8\pi G} \frac{1+8\gamma}{1+6\gamma}
\label{sigma0} 
\end{equation}
where $G = 6.67 \times 10^{-8}$ N cm$^3$ g$^{-1}$ s$^{-2}$. 
The above equation assumes that $\sigma$ is effectively massless on Solar System scales. Note that the scalar field $\sigma$ is  
effectively massless in the radiation dominated era, as can be seen in Eq.~(\ref{kg-ig}). 

The evolution of the background cosmology can be easily compared with dark energy 
in Einstein gravity with a Newton's constant $\tilde G_N$ given by the 
scalar field value at present $\tilde{G} = (8 \pi \gamma \sigma_0)^{-1}$. The Friedmann equation can be therefore rewritten by introducing an 
effective dark energy component \cite{Gannouji:2006jm}, whose energy and pressure densities for this model are \cite{FTV}:
\begin{subequations}
 \begin{align}
&\rho_{\rm DE} =\frac{\sigma_0^2}{\sigma^2}\Bigl(\frac{\dot{\sigma}}{2} - 
6\gamma H \dot{\sigma}\sigma + \lambda \frac{\sigma^4}{4} \Bigr) + \sum_i
\rho_i \left( \frac{\sigma_0^2}{\sigma^2} - 1 \right)  \nonumber \\
&p_{\rm DE}= \frac{\sigma_0^2}{\sigma^2} \Bigl[\frac{\dot{\sigma}}{2}-2\gamma H 
\dot{\sigma}\sigma -  \lambda \frac{\sigma^4}{4}  +\sum_i
\frac{2\gamma \rho_i +p_i}{1+6\gamma} \Bigr] - \sum_i p_i \,.
\end{align}
\end{subequations}
In the central panel of Fig.~(\ref{phiws}) we display the time evolution of the density contrasts of radiation 
- $\Omega_\mathrm{R} \equiv (\rho_\nu + \rho_\gamma)/(3 \gamma \sigma_0^2 H^2)$ -, matter 
 - $\Omega_\mathrm{M} \equiv (\rho_b + \rho_\mathrm{CDM})/(3 \gamma \sigma_0^2 H^2)$ - 
and effective dark energy - $\Omega_\mathrm{DE} \equiv \rho_{\rm DE}/(3 \gamma \sigma_0^2 H^2)$) -. 
As a third panel in Fig.~(\ref{phiws}), we display the time evolution of 
the parameter of state of the effective dark energy component, $w_\mathrm{DE} \equiv p_{DE}/\rho_{DE}$.

\section{The evolution of cosmological fluctuations}
\label{sec:perturbations}

As for the background, we study linear fluctuations in the Jordan frame. 
We consider metric fluctuation in the longitudinal gauge:
\begin{equation}
ds^2 = - dt^2 (1 + 2 \Psi (t, {\bf x})) + a^2(t) (1 - 2 \Phi (t, {\bf x})) dx^i dx_i
\end{equation}
and for the scalar field:
\begin{equation}
\sigma (t, {\bf x}) = \sigma(t) + \delta \sigma (t, {\bf x}) \,.
\end{equation}
The perturbed Einstein equations for our IG model with a quartic potential in the longitudinal gauge are:
\begin{subequations}
\begin{align}
 \begin{split}
  3H(\dot{\Phi}+& H\Psi) + \frac{k^2}{a^2} \Phi +3\frac{\dot{\sigma}}{\sigma}(\dot{\Phi} + 2H\Psi) - 
\frac{{\dot{\sigma}}^2}{2\gamma \sigma^2}\Psi= \nonumber \\
  - \frac{1}{2\gamma \sigma^2} \Bigl [ &3\dot{\sigma}\,\delta \dot{\sigma} - 6H^2\gamma \sigma \, 
\delta \sigma- 6H\gamma (\dot{\sigma} \, \delta \sigma+\sigma \, \dot{\delta \sigma})
- \frac{2\gamma k^2}{a^2}\delta\sigma +  \sum_i \delta \rho_i + \lambda \sigma^3 \delta \sigma \Bigr],\nonumber
 \end{split}\\
 \\[2pt]
&\dot{\Phi} +\Psi \Bigl(H+\frac{\dot{\sigma}}{\sigma}\Bigr) 
=  \frac{a}{2k^2}\frac{\sum_i (\rho_i +p_i)\theta_i}{\gamma \sigma^2}  
+\frac{\delta \sigma}{\sigma} \Bigl[ \Bigl( 1+\frac{1}{2\gamma} \Bigr)\frac{\dot{\sigma}}{\sigma} -H \Bigr] 
+\frac{\delta \dot{\sigma}}{\sigma}, \\
 &\Phi -\Psi=\frac{2\delta \sigma}{\sigma}+\frac{3a^2}{2k^2}\frac{\sum_i (\rho_i + p_i)\bar{\sigma_i}}{\gamma \sigma^2} \,.
\end{align}
\end{subequations}
In the above $\rho_i \,, p_i$ ($\delta \rho_i \,, \delta p_i$) denote the energy and (longitudinal) pressure density perturbations for each matter 
component, respectively. The velocity potential and the anisotropic stress are denoted by $\theta_i$ and $\bar{\sigma_i}$. 
We refer to Ref.~\cite{Ma:1995ey} for the conservation of the CDM, baryons, photons and neutrino 
energy-momentum tensors, since these equations are unchanged from those in Einstein gravity.  


The Klein-Gordon equation at linear order in the longitudinal gauge is:

\begin{equation}
 \begin{split}
  \delta\ddot{ \sigma} + \delta\dot{ \sigma} \Bigl( 3H +2 \frac{\dot{\sigma}}{\sigma} \Bigr) +
\Bigl[\frac{k^2}{a^2} - \frac{\dot{\sigma}^2}{\sigma^2}+\frac{\sum_i(\rho_i - 3p_i)}{(1+6\gamma)\sigma^2} \Bigr] 
\delta \sigma  \\
 =\frac{2\Psi \sum_i(\rho_i - 3p_i)}{(1+6\gamma)\sigma} +\frac{\sum_i(\delta \rho_i -3\delta p_i)}{(1+6\gamma)\sigma}
+\dot{\sigma} \Bigl(3 \dot \Phi + \dot \Psi \Bigr) \,.
 \end{split}
\end{equation}

It is interesting to note that the equation for the field fluctuation does not depend on the potential explicitly 
for the self-interacting case, as for the background in Eq.~(\ref{kg-ig}).  

We have modified the publicly available Einstein-Boltzmann code CLASS \footnote{\href{www.class-code.net}{www.class-code.net}} \cite{Lesgourgues:2011re,Blas:2011rf} to evolve background 
and linear fluctuations within induced gravity. Previous implementations of induced gravity 
in Einstein-Boltzmann codes include Refs.~\cite{Chen:1999qh,Perrotta:1999am,Riazuelo:2001mg,Nagata:2002tm,Wu:2009za}.

We initialize the fluctuation of the metric and of the matter components with adiabatic
initial condition deep in the radiation era.
We have tested our numerical results from our modified code against analytic approximations derived 
within the matter era.
To this purpose we consider the quantity $\mu (k,a)$ which parametrize the deviations of $\Psi$  
from Einstein gravity \cite{Zhao:2008bn,Zhao:2011te}. We consider the 
definition for $\mu (k,a)$ which holds also during the radiation dominated regime 
as in Ref.~\cite{Hojjati:2011ix}:
\begin{equation}
k^2 \Psi = - 4 \pi G a^2 \mu (k,a) \left[ \Delta + 3 (\rho+p) \bar{\sigma} \right]
\label{mu2}
\end{equation}
where $\Delta= \sum_i \delta \rho_i + 3aH(\rho_i +p_i) \theta_i/k^2 $, with $\delta_i=
\delta \rho_i/\rho_i$ and $\theta_i$ is the velocity potential.
Analogously we consider the deviations from Einstein gravity of the difference 
between the Newtonian potentials, parametrized by $\delta$, whose definition valid also in the 
radiation dominated regime is \cite{Hojjati:2011ix}:
\begin{equation}
k^2 [\Phi - \delta(k,a)\Psi] =  12 \pi G a^2 \mu(k,a) (\rho+p) \bar{\sigma} 
\label{gamma}
\end{equation}

\begin{figure}[t!!]
\centering
\epsfig{file=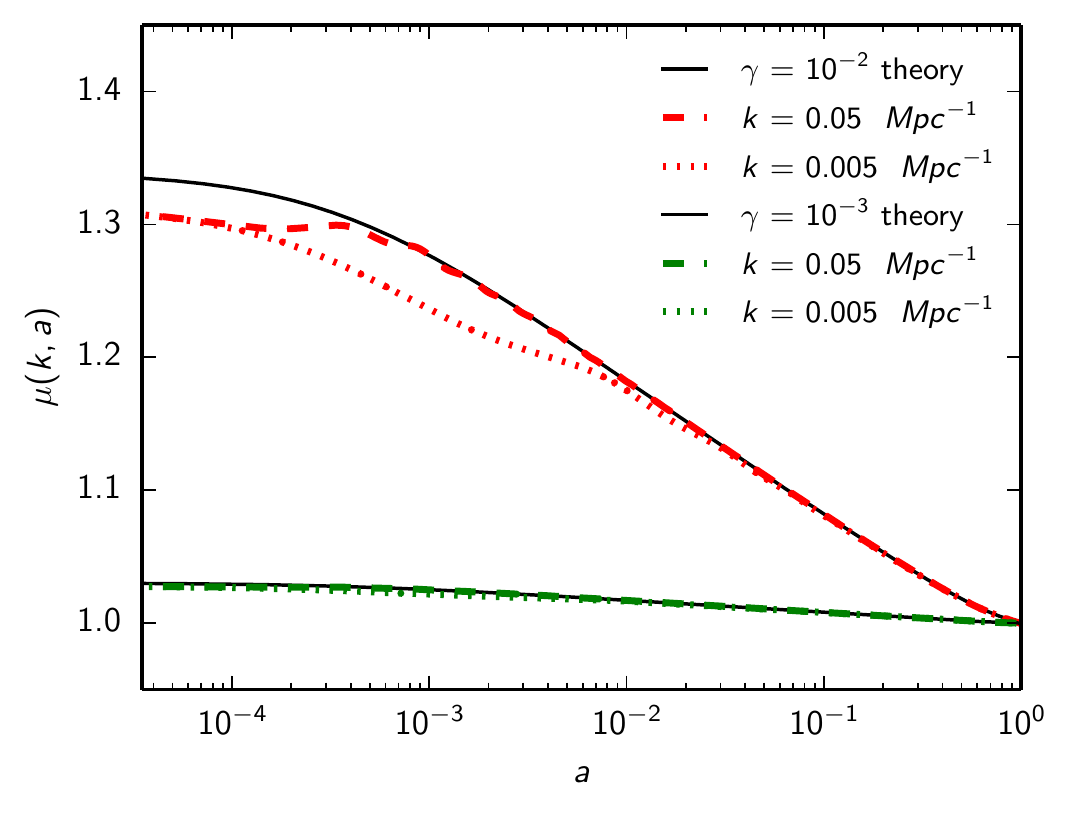, width=7.5 cm}
\epsfig{file=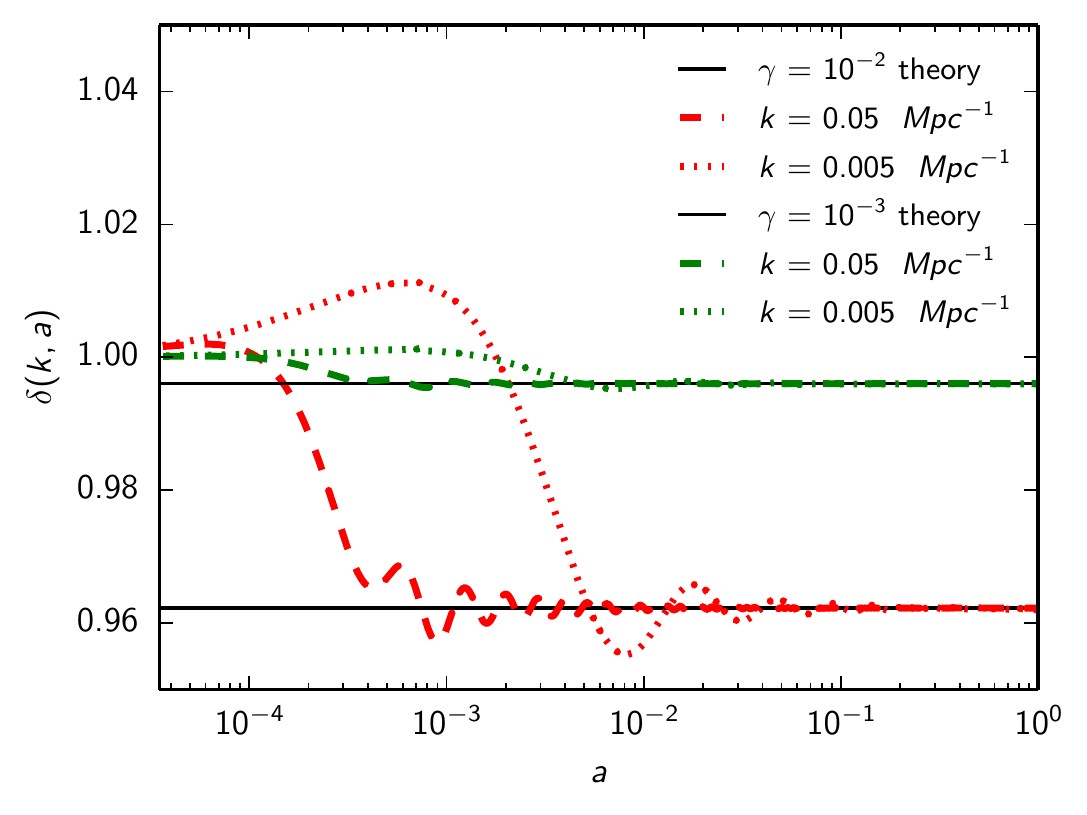, width=7.7 cm}
\caption{Comparison of theoretical approximations for $\mu$ and $\delta$ which 
parametrize deviations from Einstein gravity (black lines) with our numerical results for two 
wavenumbers ($k\ \mathrm{Mpc} = 0.05 \,, 0.005$) 
and two values of the coupling to the Ricci curvature ($\gamma=10^{-2} \,, 10^{-3}$).}
\label{mu}
\end{figure}

Our results are shown in Fig.~(\ref{mu}). 
In the left panel we show the evolution of $\mu(k,a)$ for two 
wavenumbers ($k = 0.05 \,, 0.005$ Mpc$^{-1}$)
and two values of the coupling to the Ricci curvature ({\bf $\gamma=10^{-2} \,, 10^{-3}$}). 
We compare our numerical results for $\mu (k,a)$ to the analytic approximation in the matter era: 
\begin{equation}
\mu (k,a) = \frac{\sigma_0^2}{\sigma^2}
\end{equation}
which is derived from Ref.~\cite{Zhao:2011te} 
for our choice of the potential and for our 
identification of the gravitational constant in Eq.~(\ref{sigma0}). 
Well after matter-radiation equivalence, the quasi-static analytic approximation for $k \gg a H$ in the matter era for $\mu (k,a)$ independent 
on $k$ is well recovered.
In the right panel of Fig.~(\ref{mu}), we compare the evolution of $\delta(k,a)$ for the same two wavenumbers and two values of 
the coupling $\gamma$ with the quasi-static approximation \cite{Boisseau:2000pr,Amendola:2007rr,Tsujikawa:2007gd}:
\begin{equation}
\delta (k,a) = \frac{1 + 4 \gamma}{1 + 8 \gamma} \,.
\end{equation}
Again, the analytic quasi-static approximation holds well after 
matter-radiation equality for sub-Hubble scales.

The agreement between our numerical treatment and the quasi-static approximation 
means that our self-consistent treatment of background and linear perturbations 
is sufficiently ready for precision cosmology.
The two panels in Fig.~(\ref{mu}) also show how the time evolution for $\mu$ and $\delta$, independent on $k$, 
recovered within the quasi-static approximation is not valid when the wavelength is larger 
than the Hubble radius: 
predictions for CMB anisotropies in this model would be therefore affected by considering $\delta$ constant and equal to the value obtained within the quasi-static approximation 
at {\em all} times.


\section{CMB anisotropies and Matter Power spectrum}
\label{sec:cmb}

\begin{figure}[t!!]
\centering
\epsfig{file=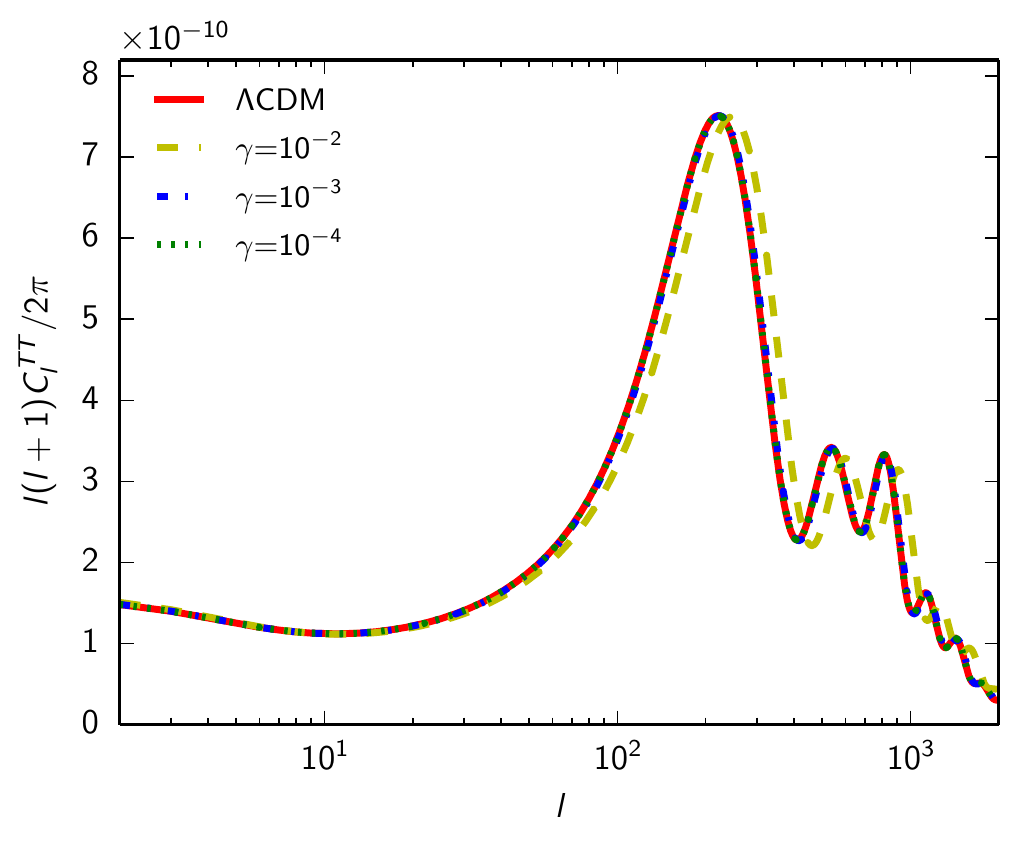, width=7.4 cm} \epsfig{file=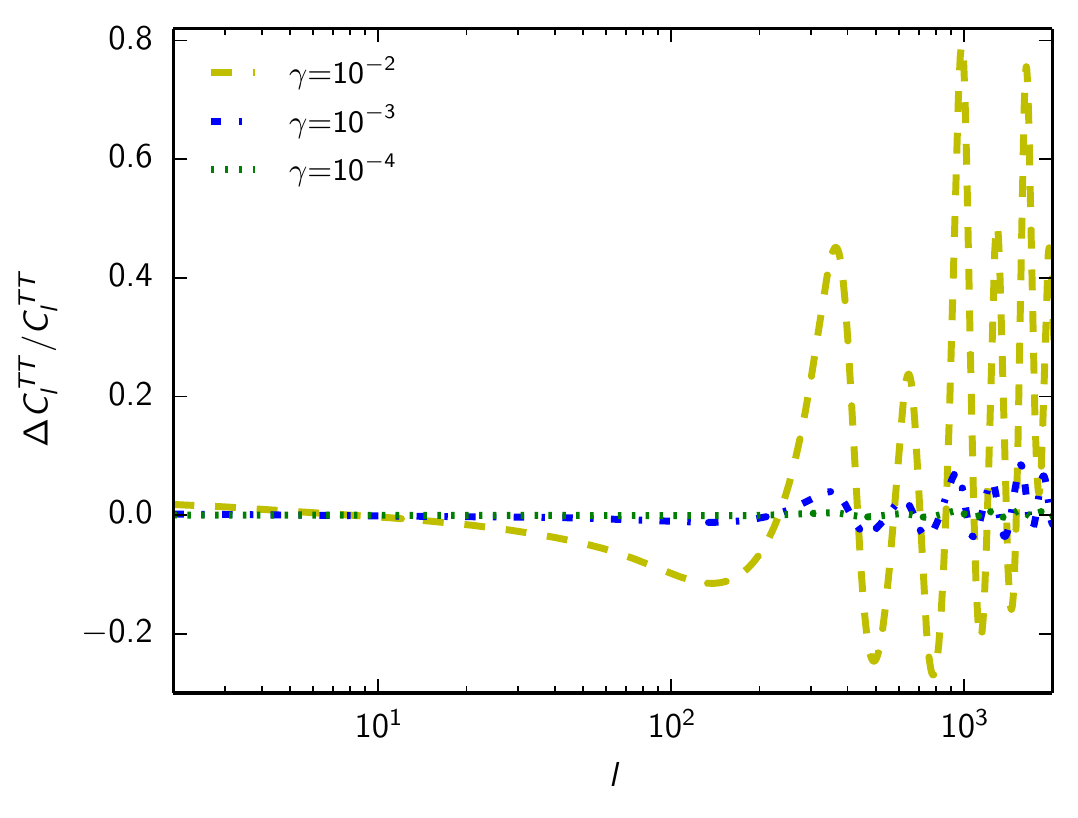, width=7.7 cm} \\
\caption{CMB temperature anisotropies power spectrum 
for different values of $\gamma$ (left panel) and relative 
differences with respect to a reference $\Lambda$CDM (right panel).}
\label{clT_delta}
\end{figure}

In the left panel of Fig.~(\ref{clT_delta}) are shown the power spectra of the CMB temperature anisotropies for different values of $\gamma$ 
($10^{-2},\ 10^{-3},\ 10^{-4}$). The relative differences with respect to the $\Lambda$CDM reference 
model are shown in the right panel of Fig.~(\ref{clT_delta}). The change in the matter-radiation 
equality present in this scalar-tensor model \cite{Liddle:1998ij} induces relative differences in the 
temperature power spectrum at few percent level for $\gamma = 10^{-3}$.

\begin{figure}[t!!]
\centering
\epsfig{file=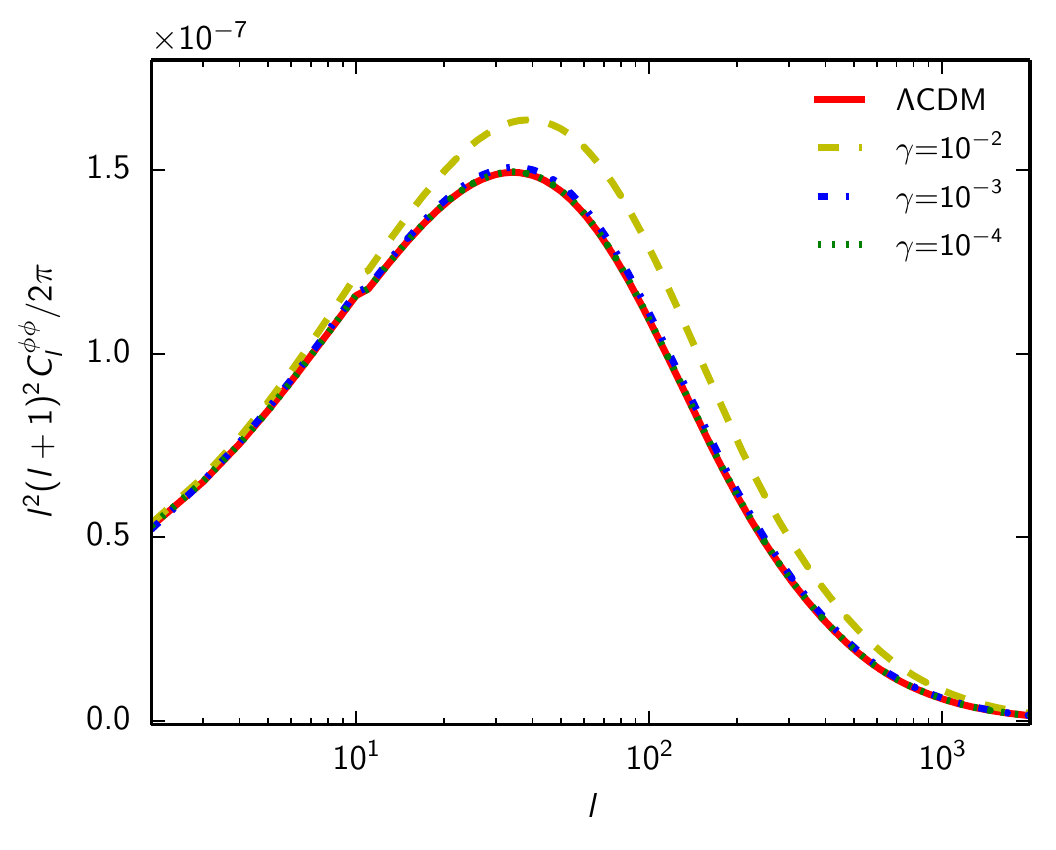, width=7.5 cm} \epsfig{file=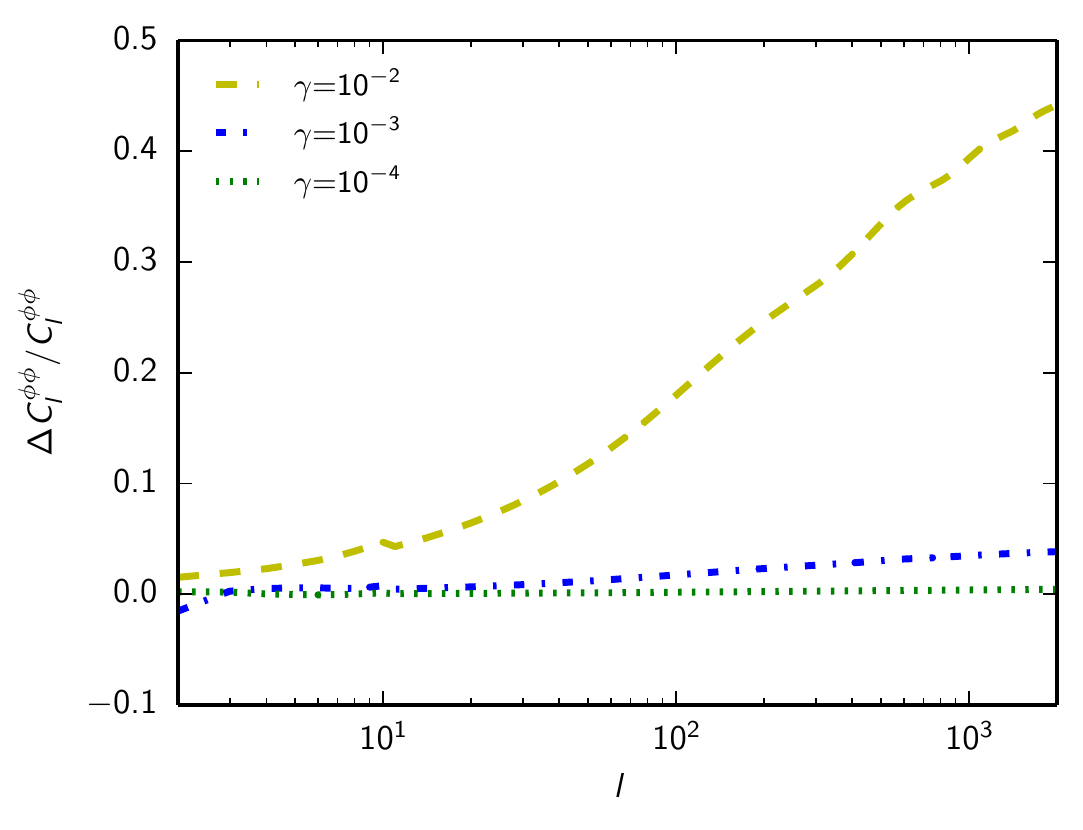, width=7.6 cm} \\
\caption{Lensing power spectrum for different values of $\gamma$ (left panel) and relative differences 
with respect to a reference $\Lambda$CDM (right panel).}
\label{clphi}
\end{figure}

\begin{figure}[t!!]
\centering
\epsfig{file=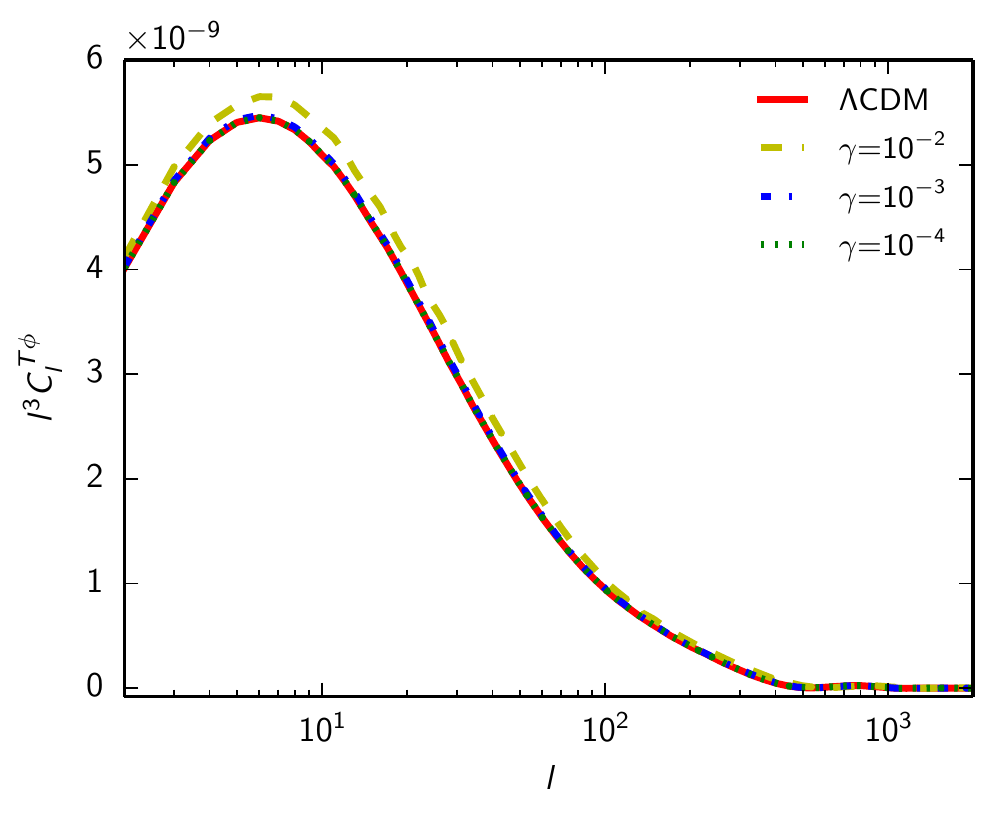, width=7.5 cm} \epsfig{file=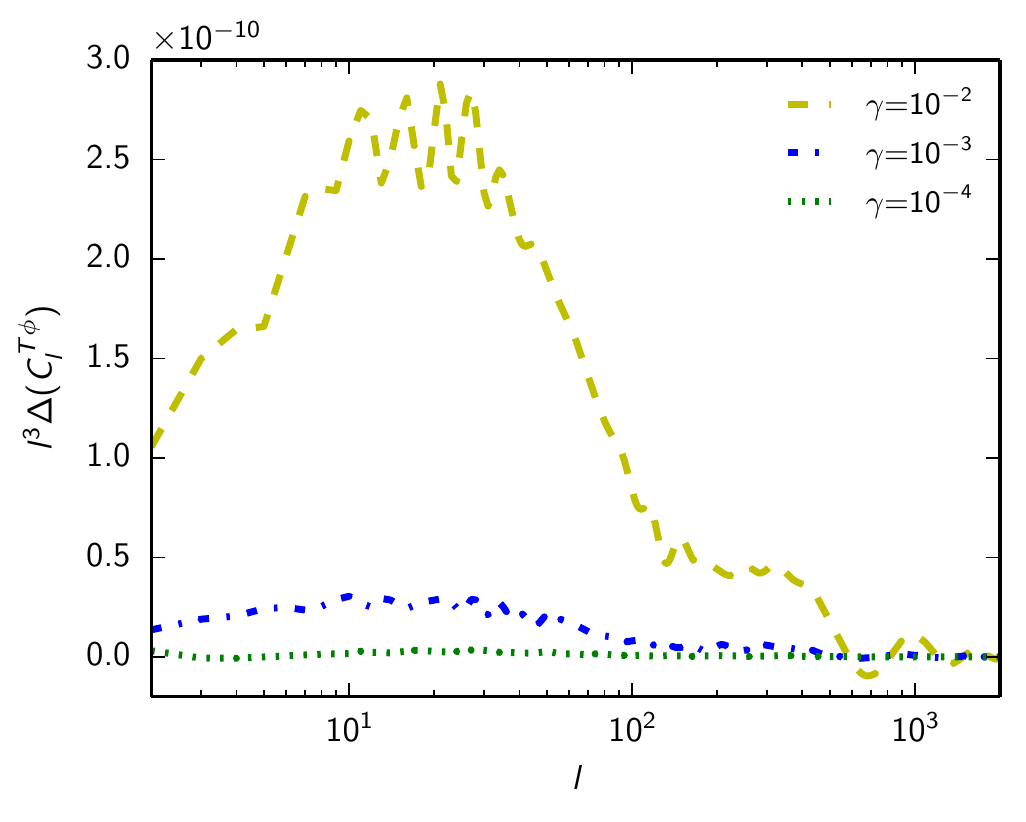, width=7.6 cm} \\
\caption{Temperature-lensing cross-correlation power spectrum for different values of $\gamma$ (left panel) and differences 
with respect to a reference $\Lambda$CDM (right panel).}
\label{clTphi}
\end{figure}

\begin{figure}[t!!]
\centering
\epsfig{file=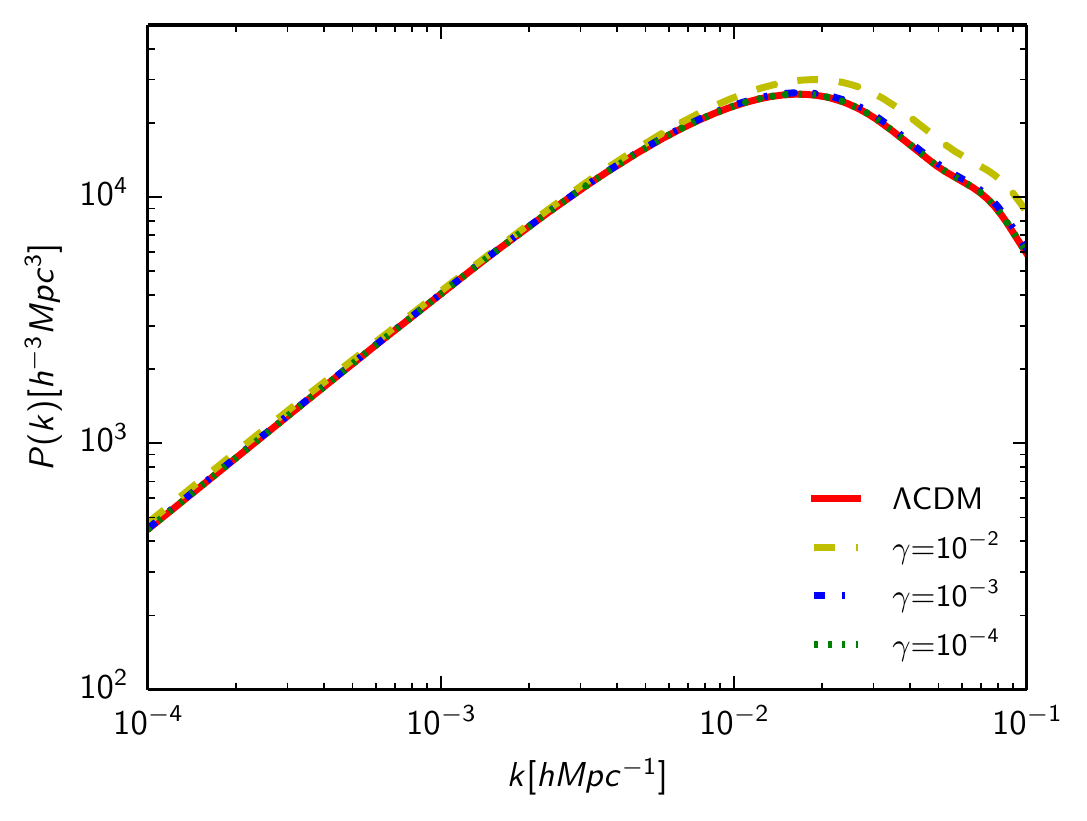, width=7.5 cm} \epsfig{file=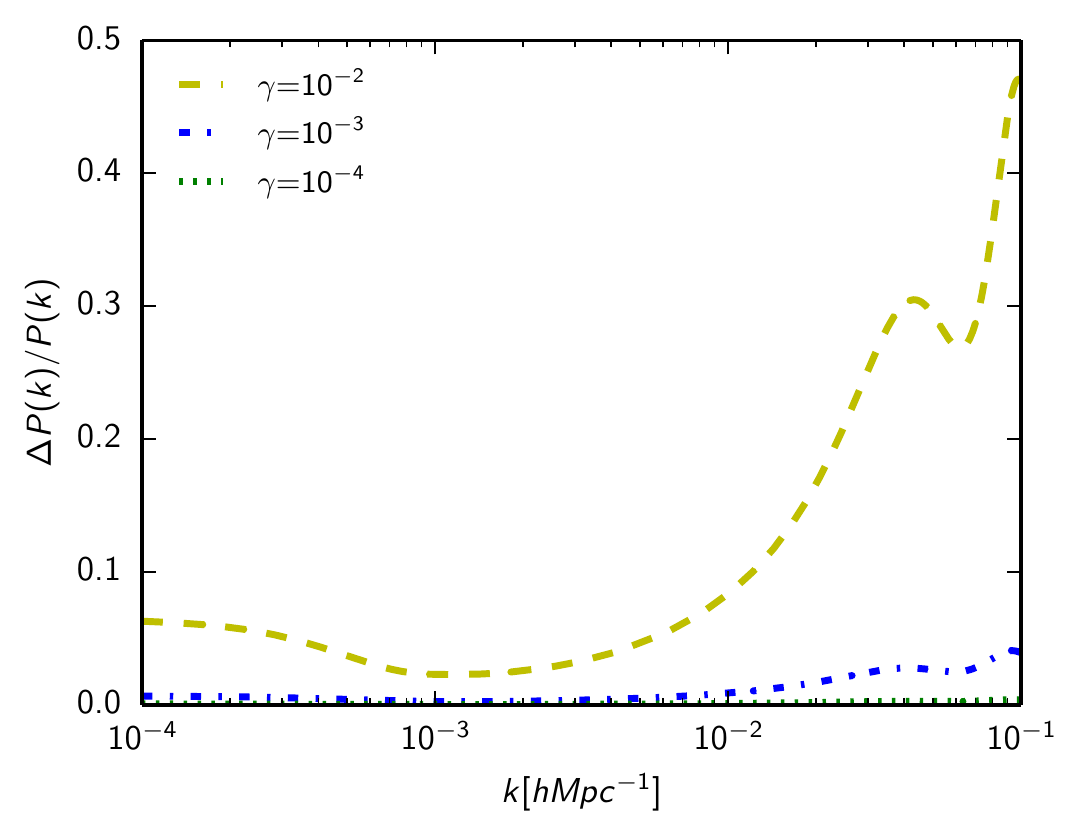, width=7.5 cm} \\
\caption{Linear matter power spectrum (at $z=0$) for different values of $\gamma$ (left panel) and relative
differences with respect to a reference $\Lambda$CDM (right panel).}
\label{matter_ps_delta}
\end{figure}

In Figs.~(\ref{clphi}, \ref{clTphi}) we display the predictions for the spectrum 
of lensing potential and its correlation with the temperature field. In Fig.~(\ref{matter_ps_delta}) 
we display the (linear) matter power spectrum at $z=0$ and the relative differences with 
respect to the $\Lambda$CDM reference model. Overall, differences at the percent level are 
obtained for $\gamma=10^{-3}$ in different cosmological observables.


\begin{figure}[t!!]
\centering
\epsfig{file=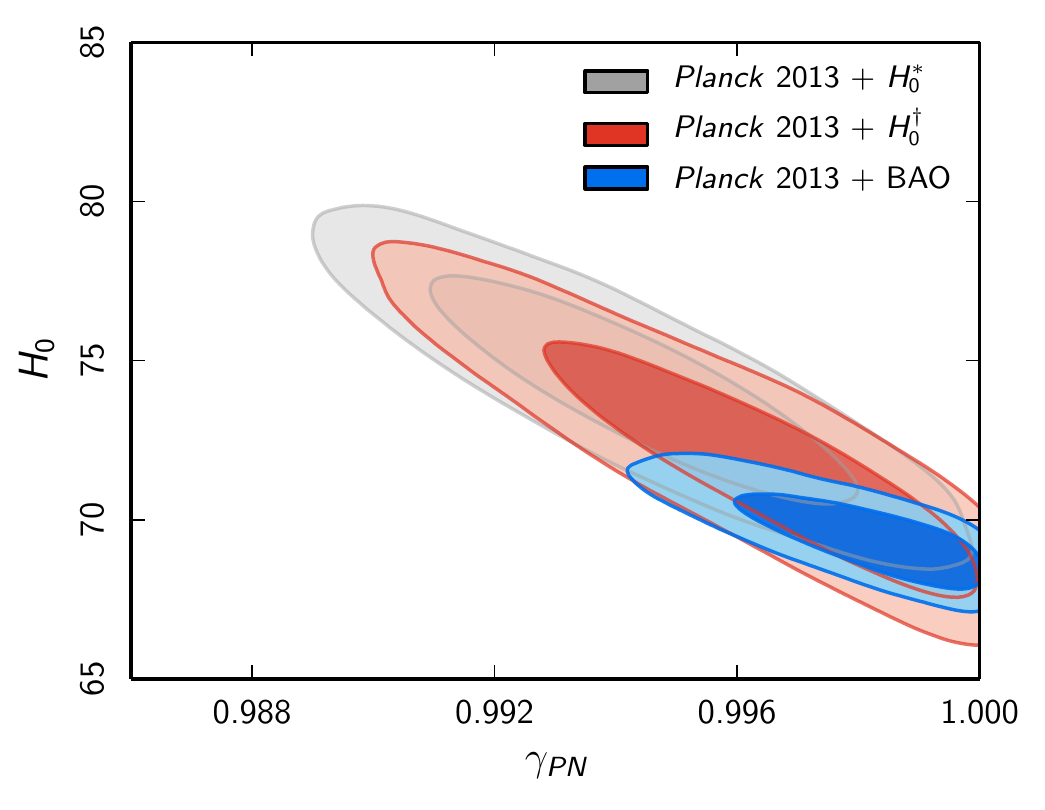, width=7.5 cm}
\epsfig{file=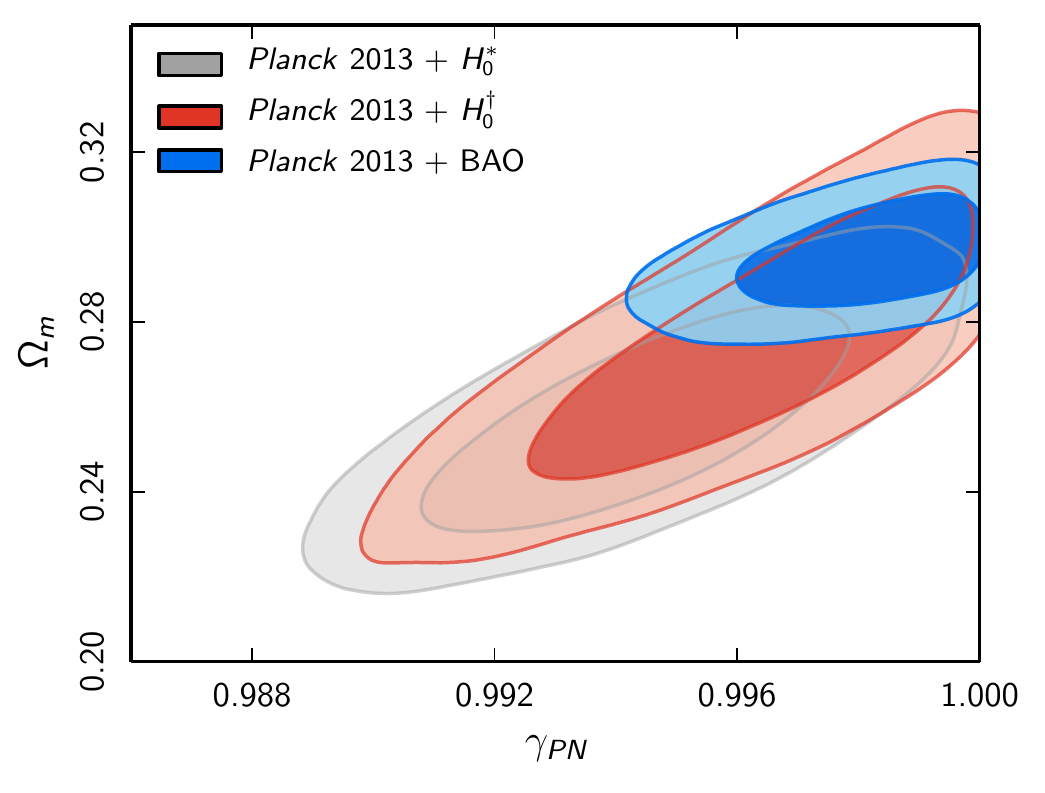, width=7.5 cm}
\caption{Comparison of marginalized joint 68\% and 95\% CL for ($\gamma_{PN},H_0$) (left panel) 
and ($\gamma_{PN},\Omega_m$) (right panel) for {\sc Planck} 2013 + $H^*_0$ (grey contours), 
{\sc Planck} 2013 + $H^\dagger_0$ (red contours) and {\sc Planck} 2013 + BAO (blue contours).}
\label{2D}
\end{figure}

\section{Constraints from cosmological observations}
\label{sec:cosmoresults}

We explore the parameter space by the Monte Carlo code for Cosmological Parameter extraction 
{\sc Monte Python}\footnote{\href{www.montepython.net}{www.montepython.net}}~\cite{Audren:2012wb} connected to the modified version 
of the Einstein-Boltzmann code CLASS used in the previous sections. 
We use the nominal mission data release from {\sc Planck}, available from the Planck Legacy Archive 
\footnote{\href{pla.esac.esa.int/pla/aio/planckProducts.html}{pla.esac.esa.int/pla/aio/planckProducts.html}}~\cite{Ade:2013ktc}. 
The {\sc Planck} likelihood covering temperature anisotropies from $\ell=2$ to $2500$ is 
combined with the low-$\ell$ WMAP polarization data \citep{Page:2006hz,bennett2012,Hinshaw:2012aka} 
(this combination is denoted as {\bf {\sc Planck}} 2013 in the following).

We use the {\sc Planck} 2013 likelihood in combination with constraints 
on $D_V(\bar{z})/r_\mathrm{s}$ (the ratio between the spherically averaged
distance scale  $D_V$ to the effective survey redshift, $\bar{z}$, and the sound
horizon, $r_\mathrm{s}$) inferred from a compilation of BAO data. 
These are 6dFGRS data \citep{Beutler:2011hx}
at $\bar{z} = 0.106$, the SDSS-MGS data \citep{Ross:2014qpa} at $\bar{z} = 0.15$,
and the SDSS-DR11 CMASS and LOWZ data \citep{Anderson:2013zyy} at
redshifts $\bar{z} = 0.57$ and $0.32$.

We vary the parameters of the flat $\Lambda$CDM model, i.e. the baryon density ($\Omega_b h^2$), the CDM density 
($\Omega_\mathrm{CDM}$), the reduced Hubble parameter ($h = H_0/(100 \mathrm{\ km\ s}^{-1} \mathrm{Mpc}^{-1})$), 
the reionization optical depth $\tau$, the amplitude and tilt of the primordial spectrum of 
curvature perturbations ($A_\mathrm{s}$ and $n_\mathrm{s}$) at the pivot scale $k_*=0.05$ Mpc$^{-1}$. 
The IG dark energy model with quartic potential is described by these six plus {\em one extra} parameter which quantifies the coupling to the 
Ricci curvature\footnote{The parameter of the Lagrangian $\lambda$ and the initial value of the 
scalar field $\sigma_i$ deep inside the radiation era are chosen to reproduce the present value of $h$ and of the field in Eq.~(\ref{sigma0}) by evolving the Friedmann and Klein-Gordon background equations.}. 
Following Ref.~\cite{Li:2013nwa} we sample on the quantity $\zeta$, defined as:
\begin{equation}
\zeta \equiv \ln \left( 1 + 4 \gamma \right) = \ln \left( 1 + \frac{1}{\omega_\mathrm{BD}}  \right)
\end{equation}
with the prior $[0,0.039]$ used in Ref.~\cite{Li:2013nwa}.
In this paper we consider three massless neutrinos \footnote{Note that the {\sc Planck} collaboration 
assumes one massive neutrinos with a mass of 0.06~eV~\cite{Ade:2013zuv}. Given the interest in 
neutrino masses within modified gravity (see for example \cite{Motohashi:2012wc}), we will study this issue in the context of induced gravity 
in a separate publication. Even if the assumption of a mass of 0.06~eV has a small effect on the cosmological 
parameters at the {\sc Planck} precision (as a $0.5 \, \sigma$ shift to smaller value for $H_0$ \cite{Ade:2013zuv}), 
we quote the results for a $\Lambda$CDM cosmology with three massless neutrinos in the following 
for a consistent comparison with the class of dark energy models studied here.}. 
Nuisance parameters for foreground, calibration and beam uncertainties \cite{Ade:2013ktc,Ade:2013zuv}.


Our results with {\sc Planck} 2013 + BAO data 
for the main and derived parameters are summarized and compared with the  
$\Lambda$CDM values in Table~\ref{tab:PlanckplusBAO}. The induced gravity model with a quartic potential is not preferred 
over Einstein gravity with $\Lambda$ {\bf ($\Delta \chi^2 \simeq - 2 \ln {\cal L} = 0.7$)}.

We quote the following {\sc Planck} 2013 + BAO 95\% CL constraint on the coupling 
to the Ricci curvature:
\be
\gamma < 0.0012 \, (95\,\%\ \text{CL, {\sc Planck} 2013 + BAO}) \,.
\label{gamma_bound}
\ee 
We quote as a derived parameter 
the corresponding constraint on the post-Newtonian parameter $\gamma_\mathrm{PN} = (1 + 4 \gamma)/(1+8\gamma)$ 
\footnote{In this class of models $\beta_\mathrm{PN}=1$.}:
\be
0.9953 < \gamma_\mathrm{PN} < 1 \, \, (95\,\%\ \text{CL, {\sc Planck} 2013 + BAO}) \,.
\label{gammaPN_bound}
\ee

It is also useful to quote the derived constraints on the change of the Newton constant between the radiation era and the present time 
$\delta G_N/G_N \equiv (\sigma_i^2 - \sigma^2_0)/\sigma_0^2$:
\be
\frac{\delta G_N}{G_N} = -0.015_{-0.006}^{+0.013} \, (95\,\%\ \text{CL, {\sc Planck} 2013 + BAO})
\label{deltaG_bound}
\ee
and the constraint on its derivative ($\dot G_N/G_N \equiv - 2 \dot \sigma_0 / \sigma_0$) at present time:
\be
\frac{\dot{G}_N}{G_N} (z=0) = -0.61_{-0.25}^{+0.55} \, [\times10^{-13}\ yr^{-1}] \,, (95\,\%\ \text{CL, {\sc Planck} 2013 + BAO}) \,.
\label{dotG_bound}
\ee

The constraints derived here are tighter than those obtained in the literature with 
{\sc Planck} 2013 data for similar scalar-tensor  
models with a power-law potential \citep{Avilez:2013dxa,Li:2013nwa} (see Refs.~\cite{Nagata:2003qn,Acquaviva:2004ti,Wu:2009zb} 
for analysis with pre-{\sc Planck} data). 
Avilez and Skordis \cite{Avilez:2013dxa} considered 
the case of a constant potential in Brans-Dicke-like theory 
and quote $(1 + 6 \gamma)/(1+8 \gamma) = 1.07_{-0.10}^{+0.11}$ at 95\% CL as the tightest 
constraint with a prior $\omega_\mathrm{BD} > -3/2$; we obtain [0.998,1] as the 95\% CL range for the same quantity with {\sc Planck} 2013 + BAO 
by varying $\zeta$ in the interval $[0,0.039]$. 
Li et al. \cite{Li:2013nwa} considered the case of a linear potential in Brans-Dicke 
(i.e., a quadratic potential in induced gravity) and quote $0 < \zeta < 0.549 \times 10^{-2}$ at 95\% CL and $\dot G_N/G_N = - 1.42^{+2.48}_{-2.27}$ at 
68\% CL from {\sc Planck} 2013 with the same prior 
on $\zeta$, although in combination with a different compilation of BAO data \citep{Li:2013nwa}. 
Note that for power-law potentials different from the quartic case studied here, Einstein gravity plus a cosmological constant 
with $\sigma$ independent on time is not the attractor at future times \cite{CFTV}. We therefore expect that the models studied 
in Refs.~\cite{Avilez:2013dxa,Li:2013nwa} differ from the case of a quartic potential, in particular at recent redshifts.





\begin{table*}
\centering
\begin{tabular}{l|cc}
\hline
\hline
                                              & {\sc Planck} 2013 + BAO       & {\sc Planck} 2013 + BAO     \\
                                              & $\Lambda$CDM \, \,            &                             \\
\hline
$ 10^5\Omega_\mathrm{b}h^2$                   & $2215^{+24}_{-25}$           & $2203\pm25$         \\
$ 10^4\Omega_\mathrm{c}h^2$                   & $1187^{+13}_{-14}$           & $1207_{-22}^{+18}$         \\
$H_0$ [km s$^{-1}$ Mpc$^{-1}$]                & $68.4^{+0.6}_{-0.7}$         & $69.5_{-1.2}^{+0.9}$    \\
$\tau$                                        & $0.091^{+0.012}_{-0.014}$    & $0.088_{-0.013}^{+0.012}$  \\
$\ln \left(  10^{10} A_\mathrm{s} \right)$    & $3.089^{+0.024}_{-0.027}$    & $3.090_{-0.026}^{+0.024}$  \\
$n_{\mathrm s}$                               & $0.9626\pm0.0053$            & $0.9611 \pm 0.0053$        \\
$\zeta$                                       & $...$                        & $<0.0047$ (95\% CL)        \\
\hline
$10^3 \gamma$                                 & $...$                          & $<1.2$ (95\% CL)          \\
$\gamma_{PN}$                                 & $...$                          & $>0.9953$ (95\% CL)        \\
$\Omega_\mathrm{m}$                           & $0.301\pm0.008$                          & $0.295\pm0.009$                        \\
$\delta G_\mathrm{N}/G_\mathrm{N}$            & $...$                          & $-0.015_{-0.006}^{+0.013}$ \\
$10^{13} \dot{G}_\mathrm{N}(z=0)/G_\mathrm{N}$ [yr$^{-1}$]   & $...$                          & $-0.61_{-0.25}^{+0.55}$    \\
$10^{23} \ddot{G}_\mathrm{N}(z=0)/G_\mathrm{N}$ [yr$^{-2}$]  & $...$                       & $0.86_{-0.78}^{+0.33}$     \\
\hline
\hline
\end{tabular}
\caption{\label{tab:PlanckplusBAO} 
Constraints on main and derived parameters (at 68\% CL if not otherwise stated).
}
\end{table*}

\subsection{Combination with local measurements}

As from Table~\ref{tab:PlanckplusBAO}, the model considered here prefers a higher value of the Hubble parameter $H_0$ with respect to $\Lambda$CDM. 
We therefore analyze the combination of the local measurements of the Hubble constant with {\sc Planck} 2013 and BAO data.
The local estimates of $H_0$ are consistently higher than those from CMB (and BAO) and this discrepancy became 
a 2.5 {\bf $\sigma$} tension after the {\sc Planck} 2013 release \cite{Ade:2013zuv}. 
This tension might be sign of new physics, although reanalysis subsequent to the {\sc Planck} 2013 release 
have highlighted how hidden systematics and underestimated uncertainties could hide in the local measurements of $H_0$ 
\cite{Humphreys:2013eja,Efstathiou:2013via}.
For these reasons we consider separately the impact of two different local estimates of $H_0$, such as 
$H_0 = 73.8 \pm 2.4$ km s$^{-1}$ Mpc$^{-1}$ \cite{Riess:2011yx}, denoted as $H_0^*$, and 
$H_0 = 70.6 \pm 3.0$ km s$^{-1}$ Mpc$^{-1}$ \cite{Efstathiou:2013via}, denoted as $H_0^\dagger$.
Our results are summarized in Table~\ref{tab:HZ} and Fig.~(\ref{2D}). 

With the higher local estimate of $H_0^*$ \cite{Riess:2011yx} we obtain a posterior on $\zeta$ which 
is different at 2 $\sigma$ level from Einstein gravity. With the lower estimate for $H_0$ 
obtained by Efstahiou \cite{Efstathiou:2013via} with the new revised geometric maser 
distance to NGC 4258 \cite{Humphreys:2013eja}, we obtain a posterior probability for $\zeta$ compatible with Einstein gravity.  
As can be seen by comparing the last columns of Table~\ref{tab:PlanckplusBAO} and \ref{tab:HZ}, 
the lower local estimate of $H_0^\dagger$ has almost a negligible impact when {\sc Planck} 2013 and BAO data are combined.

\begin{table*}
\centering
\begin{tabular}{l|ccc}
\hline
\hline
                                              & {\sc Planck} 2013 + $H_0^*$         & {\sc Planck} 2013 + $H_0^\dagger$ & {\sc Planck} 2013 + BAO + $H_0^\dagger$      \\
\hline
$ 10^5\Omega_\mathrm{b}h^2$                   & $2219 \pm 28$                  & $2213_{-29}^{+28}$ 			  & $2203 \pm 26$                \\
$ 10^4\Omega_\mathrm{c}h^2$                   & $1188_{-26}^{+25}$             & $1194_{-25}^{+25}$ 			  & $1207_{-22}^{+18}$           \\
$H_0$ (km s$^{-1}$ Mpc$^{-1}$)                & $74.1_{-2.4}^{+2.3}$           & $72.1_{-3.1}^{+2.2}$ 			  & $69.64_{-1.11}^{+0.88}$      \\
$\tau$                                        & $0.092_{-0.014}^{+0.013}$      & $0.091_{-0.015}^{+0.013}$ 		  & $0.088_{-0.014}^{+0.012}$    \\
$\ln \left(  10^{10} A_\mathrm{s} \right)$    & $3.098_{-0.027}^{+0.025}$      & $3.095_{-0.028}^{+0.025}$ 		  & $3.091_{-0.027}^{+0.024}$    \\
$n_{\mathrm s}$                               & $0.9704_{-0.0072}^{+0.0070}$   & $0.9667_{-0.0078}^{+0.0075}$		  & $0.9613_{-0.0054}^{+0.0055}$ \\
$\zeta$                                         & $0.0056 \pm 0.0023$            & $<0.0083$ (95\% CL) 			  & $0.0047$ (95\% CL)           \\
\hline
$10^3 \gamma$                                 & $1.4\pm0.6$         & $<2.1$ (95\% CL)			  & $<1.2$ (95\% CL)            \\
$\gamma_{PN}$                                 & $0.9944_{-0.0022}^{+0.0023}$     & $>0.9918$ (95\% CL)                      & $>0.9954$ (95\% CL)          \\
$\Omega_\mathrm{m}$                           & $0.257_{-0.019}^{+0.016}$      & $0.274_{-0.021}^{+0.022}$                & $0.294_{-0.008}^{+0.009}$                          \\
$\delta G_\mathrm{N}/G_\mathrm{N}$            & $-0.041_{-0.016}^{+0.017} $    & $-0.028 \pm 0.012$			  & $-0.016_{-0.006}^{+0.010}$   \\
$10^{13} \dot{G}_\mathrm{N}(z=0)/G_\mathrm{N}$  [yr$^{-1}$]   & $-1.56_{-0.58}^{+0.61}$        & $-1.10_{-0.49}^{+0.83}$		  & $-0.64_{-0.25}^{+0.52}$      \\
$10^{23} \ddot{G}_\mathrm{N}(z=0)/G_\mathrm{N}$  [yr$^{-2}$] & $2.4_{-1.0}^{+0.9}$            & $1.7_{-1.5}^{+0.7}$			  & $0.89_{-0.75}^{+0.24}$       \\
\hline
\hline
\end{tabular}
\caption{\label{tab:HZ} 
Constraints on main and derived parameters at 68\% CL (if not otherwise stated). 
}
\end{table*}

\section{Conclusions}
\label{sec:concl}
We have studied structure formation in a induced gravity dark energy model with a quartic potential. 
In this model the current acceleration stage of the Universe and an accompanying change in the gravitational constant on large scales 
are due to a change in the background scalar field triggered by the onset of the matter dominated stage \cite{CV,FTV}.

We have shown that the model approaches Einstein gravity plus a cosmological 
constant in the limit $\gamma \rightarrow 0$
{\em also} at linear level. We have shown how the quasi-static parametrization with 
$\mu$ and $\gamma$ independent on $k$ holds only well after matter-radiation equality for sub-Hubble scales.

We have derived CMB and BAO combined constraints on $\gamma$ for the case of induced gravity with a quartic potential.
By using {\sc Planck} 2013 \cite{Ade:2013kta} and BAO \cite{Beutler:2011hx,Ross:2014qpa,Anderson:2013zyy} 
data we derive the 95\%CL constraint $\gamma < 0.0012$, which is tighter than previous 
cosmological constraints on similar models \cite{Avilez:2013dxa, Li:2013nwa}. This cosmological 
constraint is compatible, but weaker than those within the Solar System \cite{FTV} which can be derived by 
Cassini data \cite{Bertotti:2003rm}. Since there is a positive correlation between $\gamma$ and $H_0$, the combination of local measurements of 
$H_0$ \cite{Riess:2011yx,Efstathiou:2013via} allows larger values of $\gamma$, but not at statistical significant level. 
This analysis shows how a self-consistent variation of $G$ from the radiation era to the present time 
can be tightly constrained from {\sc Planck} 2013 and BAO data at the percent level. 
It will be interesting to see how {\sc Planck} 2015 data \cite{Adam:2015rua} 
change these constraints \cite{Ballardini2015}.


\section*{Acknowledgements}
We wish to thank Julien Lesgourgues, Thomas Tram and Benjamin Audren for many useful suggestions 
on the use of the CLASS code and MONTEPYTHON code. 
We wish to thank Massimo Rossi for careful checking of the equations for cosmological perturbations 
and Levon Pogosian for useful comments on the draft. 
This work has been done within the Labex ILP (reference
ANR-10-LABX-63) part of the Idex SUPER, and received financial state
aid managed by the Agence Nationale de la Recherche, as part of the
programme Investissements d'avenir under the reference
ANR-11-IDEX-0004-02.
The support by the "ASI/INAF Agreement 2014-024-R.0 for the Planck LFI Activity of Phase E2" is acknowledged.



\end{document}